\documentstyle[preprint,aps,epsf]{revtex}

\newcommand {\bea}{\begin{eqnarray}}
\newcommand {\eea}{\end{eqnarray}}
\newcommand {\be}{\begin{equation}}
\newcommand {\ee}{\end{equation}}

\begin{document}

\preprint{IASSNS-HEP 99/32}

\title{Quark Description of Hadronic Phases}

\author{Thomas Sch\"afer\footnote{Research supported in part by NSF
PHY-9513835.  e-mail: schaefer@sns.ias.edu}
and Frank Wilczek\footnote{Research supported in part by DOE grant
DE-FG02-90ER40542. e-mail: wilczek@sns.ias.edu }}

\address{School of Natural Sciences\\
Institute for Advanced Study\\
Princeton, NJ 08540}

\maketitle

\begin{abstract}

We extend our proposal that major universality classes of hadronic
matter can be understood, and in favorable cases calculated, directly
in the microscopic quark variables, to allow for splitting between
strange and light quark masses.  A surprisingly simple but apparently
viable
picture emerges, featuring essentially three phases,
distinguished by whether strangeness is conserved
(standard nuclear matter), conserved modulo two (hypernuclear matter),
or locked to color (color flavor locking).  These are
separated by sharp phase
transitions.  There is also, potentially, a quark phase matching
hadronic K-condensation.
The smallness of the secondary gap in two-flavor color superconductivity
corresponds to the
disparity between the primary dynamical energy scales of QCD and
the much smaller energy scales of nuclear physics.

\end{abstract}

\newpage

In principle QCD ought to be an adequate basis for the description of
hadronic matter in extreme conditions, including extremely high
densities such as might be achieved deep in neutron star interiors, or
transiently during collapse of massive stars, collisions between
neutron stars, and laboratory heavy ion collisions.  In practice
however our ability to solve the equations is quite limited, and it
has been challenging to provide a description firmly rooted in the
microscopic theory.  Specifically, the relationship between widely
used phenomenological models, based on hadron degrees of freedom, and
models based on quark degrees of freedom, themselves typically
employing highly idealized abstractions of real QCD dynamics, has been
quite unclear.  It has generally been supposed that at relatively low
density the hadronic description is appropriate, while at higher
density some sort of quark description is appropriate.

Recently we argued for the radical proposal that in a slightly
idealized version of QCD, with three degenerate flavors (and ignoring
electromagnetism) there need be no sharp distinction between these
pictures \cite{SW_98b}.  More precisely, we argued that the color-flavor
locked state \cite{ARW_98b} (which can now, thanks to \cite{Son_98}, be
pretty rigorously identified as the correct ground state at
asymptotically
large densities), though firmly based on quark and gluon degrees of
freedom, exhibits all the main features one might expect from naive
extrapolation of phenomenologically-based conventional wisdom for the
low-density phase.  Specifically, the ground state exhibits
confinement and chiral symmetry breaking, while the elementary
excitations carry quantum numbers of pseudoscalar mesons,
vector mesons, and
baryons matching ``phenomenological'' expectations, notably
including integral electric charges.  The simplest hypothesis, then,
is that there is only one phase.  This implies, in particular, that
color-flavor locking provides a rigorously defined, asymptotically
controlled framework wherein the classic qualitative barriers to relating
the microscopic (quark-gluon) to the macroscopic (hadron)
degrees of freedom in QCD have been overcome.

It is obviously of great interest to extend this picture to real QCD,
wherein the disparity between strange and non-strange masses is far
from negligible.  That is what we will sketch out here.

Our basic technique to understand QCD at high density is to
assume weak coupling tentatively, and to work out the consequences of
this assumption.  Indeed, high density implies large Fermi surfaces, and
therefore large momenta for an important class of low-energy degrees
of freedom, namely the quasiparticle excitations near the Fermi
surface.   Their generic interactions will be characterized by large
momentum transfer, and
one might anticipate that asymptotic freedom can be invoked
to analyze these interactions perturbatively.
In the normal groundstate, however, there are two kinds of
charged low-energy
degrees of freedom -- particle-hole excitations near the Fermi
surface, and elementary gluons.  These are dangerous, because
their exchanges generate infrared
divergences, which invalidate a straightforward perturbative approach.

Fortunately, we can do better.
Indeed, as we know from the theory of
superconductivity \cite{Sch_64} and of He$^3$ superfluidity
\cite{he3}, arbitrarily weak attractive interactions near the Fermi
surface drive a pairing instability.  Early applications of these
ideas to QCD are summarized in \cite{BL_84}.
In favorable cases, including
QCD with three or more degenerate flavors
\cite{SW_98b},  interaction
with the resulting condensate opens up gaps in all charged
channels.  Then there are no infrared problems, and the weak-coupling
analysis is internally consistent.  Of course this successful
weak-coupling approach to high-density QCD
is not the straightforward perturbative one: it is heavily rooted
in asymptotic freedom and BCS theory, and includes both
local and global spontaneous symmetry breaking (confinement and chiral
symmetry breaking), derived from first principles.

\section{Onsets and Mismatches}

Until further notice we shall consider the case where the chemical
potentials for all three flavors are set equal, and we shall pretend
that electromagnetism has been turned off.

The preferred color-flavor locking ground state is particularly
symmetric and energetically favorable for 3 degenerate flavors.
What
happens as we turn up the strange quark mass?

There are two main qualitative effects associated with the non-zero
strange quark mass.  Both can be identified either in a hadron picture
or in a quark picture.

The first effect
is that there are now two onset transitions.  From the point
of view of hadrons, one occurs
when the chemical potential for baryon number exceeds the minimal
energy/baryon in ordinary (nonstrange) nuclear matter, and the other
when the chemical potential is large enough that strange baryons are
also produced.

On the quark side, for free quarks, one would
likewise expect two onsets, occurring when the chemical potential
exceeds first the light and then the strange quark masses.  Taking
into account interactions, the situation is more complicated, though the
qualitative outcome is very likely the same.  The complication
is that the onset transition is expected to (and, from a
phenomenological point of view, had better) occur at a significantly
larger value of the chemical potential, characteristic of nucleon
rather than light quark masses.  In particular, it should
occur at a finite value of the chemical potential even for massless
quarks.  The possibility, for interacting massless quarks,
of a first-order phase transition from
the ``void'' state with chiral symmetry broken to a color
superconducting state at finite density was raised within 2-flavor
QCD in \cite{ARW_98}, \cite{RSSV_98}.  It is implicit in the
construction of the MIT bag model.

The second effect is that the light and strange Fermi surfaces will no
longer be of equal size. Let us discuss this in more detail on the
quark side.
When the
mismatch is much smaller than the gap one calculates
assuming degenerate quarks, we might
expect that it has very little consequence, since at this level
the original particle and hole states near the Fermi surface are mixed up
anyway.  On the other hand, when the mismatch is much
larger than the nominal gap,
we might expect that the ordering one would obtain for degenerate
quarks is disrupted, and that to a first approximation
one can treat the light and heavy quark dynamics separately.

In particular, for weak coupling and Fermi surface mismatch, one
should not expect mixed nonstrange-strange condensation.
Indeed, unlike when one had matched Fermi surfaces, in this case
one cannot form pairs of equal
and opposite momenta, that the elastic scatterings allowed in Fermi
liquid theory connect.  So there is not a large space of
degenerate
states connected by the relevant parts of the Hamiltonian, a weak
perturbation has small effects.
Thus, in the mixed channel one does not find  Cooper instabilities at
weak coupling.

One way to see this in a more quantitative fashion is to study a
a schematic gap equation that describes the spin singlet pairing of
two fermions with different masses. In a basis of particles of
the first kind and holes of the second the quadratic part of the
action is
\be
\left(\psi^\dagger_{(1)}\hspace{0.3cm} \psi_{(2)}\right)
\left(\begin{array}{cc}
p_0-\epsilon_p^1  & \Delta \\
\Delta^* & p_0+\epsilon_p^2
\end{array}\right)
\left( \begin{array}{c}
\psi_{(1)} \\ \psi^\dagger_{(2)}
\end{array}\right) .
\label{gorkov}
\ee
Here, $\epsilon^{1,2}_p=E^{1,2}_p-\mu$ and $E^{1,2}_p=\sqrt{p^2
+m_{1,2}^2}$ where $m_{1,2}$ are the masses of particle one and
two. The particle and hole propagators are determined by the
inverse of the matrix (\ref{gorkov}). The off-diagonal (anomalous)
propagator is
\be
\frac{\Delta}{(p_0-\epsilon^1_p)(p_0+\epsilon^2_p)-\Delta^2}.
\ee
We study the effect of a zero range interaction $G(\psi_1
\sigma_2\psi_2)(\psi^\dagger_1\sigma_2\psi^\dagger_2)$. The
pairing is described by the gap equation
\be
 \Delta= G\int\frac{d^4p}{(2\pi)^4}\frac{\Delta}
 {(p_0+R+i\delta {\rm sgn}(p_0))^2-\bar\epsilon_p^2-\Delta^2}.
\ee
Here, we have introduced $\bar\epsilon_p=\bar E_p-\mu=(\epsilon^1_p
+\epsilon^2_p)/2$ and $R=(\epsilon^1_p-\epsilon^2_p)/2$. In practice,
we are interested in pairing between almost massless up or down quarks
and massive strange quarks. In that case, $R\simeq m_s^2/(4p_F)\simeq
m_s^2/(4\mu)$. The poles of the anomalous propagator are located at
$p_0=-R\pm\sqrt{\bar\epsilon^2_p+\Delta^2}-i {\rm sgn}(p_0)$. As usual, we
close the integration contour in the lower half plane. Let us denote
the solution of the gap equation in the case $R=0$ by $\Delta_0$.
Then, if $R<\Delta_0$, the pole with the positive sign of the square
root is always included in the integration contour and we have
\be
 \Delta = \frac{G\mu^2}{4\pi^2} \int d\bar\epsilon_p\,
 \frac{\Delta}{\sqrt{\bar\epsilon^2_p+\Delta^2}}.
\label{gap}
\ee
This result is, up to a small correction in the density of states
that we have neglected here, identical to the gap equation for
degenerate fermions, so $\Delta \approx \Delta_0$. If, on the other hand,
$R>\Delta_0$ there only is a pole in the lower half plane if
$\bar\epsilon_p>\sqrt{R^2-\Delta^2}$. Carrying out the $p_0$
integration again leads to the gap equation (\ref{gap}), but
with the $\bar\epsilon_p$ integration restricted by the condition
just mentioned. This cuts out the infrared singularity at
$\bar\epsilon_p=0$ and one can easily verify that the gap equation
does not have a non-trivial solution for weak coupling. We 
thus conclude that a necessary condition for pairing is that 
\be
 m_s^2<4p_F\Delta(\mu).
\ee
So far, we have only dealt with a simple pair condensate involving
strange and non-strange quarks. In practice, we are interested in a
somewhat more complicated situation. In particular, we want to consider
the transition between the color-flavor locked phase for small $m_s$
and the two-flavor color superconductor in the limit of large $m_s$.
This analysis can be carried out along the same lines as the toy
model discussed above. We now consider the following free action
\be
\left(\psi^\dagger\hspace{0.3cm} \psi\right)
\left(\begin{array}{cc}
(p_0-\epsilon_p)1\!\!1 - 2RM_s & \Delta_{ud}M_{ud}+\Delta_{us}M_{us} \\
\Delta_{ud}M_{ud}+\Delta_{us}M_{us} & (p_0+\epsilon_p)1\!\!1+2RM_s
\end{array}\right)
\left( \begin{array}{c}
\psi  \\ \psi^\dagger
\end{array}\right) ,
\label{gor_cfl}
\ee
where $\psi$ is now a 9 component color-flavor spinor.
$M_s, M_{ud}$ and $M_{us}$ are color-flavor matrices
\bea
M_s    &=& \delta^{\alpha\beta}\delta_{a3}\delta_{b3}\\
M_{ud} &=& \epsilon^{3\alpha\beta}\epsilon_{3ab}, \\
M_{us} &=& \epsilon^{2\alpha\beta}\epsilon_{2ab}
          +\epsilon^{1\alpha\beta}\epsilon_{1ab},
\eea
where $\alpha,\beta$ are color, and $a,b$ flavor indices.
$\Delta_{ud}$ is the gap for $\langle ud\rangle$ condensation, and
$\Delta_{us}$ is the gap for $\langle us\rangle=\langle ds\rangle$
condensation. Color-flavor locking corresponds to the case $\Delta_{ud}
=\Delta_{us}$, and the two flavor superconductor corresponds to
$\Delta_{us}=0,\,\Delta_{ud}\neq 0$. Flavor symmetry breaking is
again caused by $R\simeq m_s^2/(4p_F)$. The $18\times 18$ matrix
(\ref{gor_cfl}) can be diagonalized exactly. The eigenvalues and their
degeneracies are
\be
\begin{array}{lc}
p_0 \pm  \left(\epsilon_p^2+\Delta_{ud}^2\right)^{1/2},  & d=3 \\
p_0 - R \pm \left(\bar\epsilon_p^2+\Delta_{us}^2\right)^{1/2}, & d=2\\
p_0 + R \pm \left(\bar\epsilon_p^2+\Delta_{us}^2\right)^{1/2}, & d=2\\
p_0 \pm \left( \epsilon_p^2 + 2R\epsilon_p +2R^2 + 2\Delta_{us}^2
   +\frac{1}{2}\Delta_{ud}^2 -\frac{1}{2}S\right)^{1/2},     & d=1\\
p_0 \pm \left( \epsilon_p^2 + 2R\epsilon_p +2R^2 + 2\Delta_{us}^2
   +\frac{1}{2}\Delta_{ud}^2 +\frac{1}{2}S\right)^{1/2},     & d=1
\end{array}
\label{evals}
\ee
where
\be
S=\left(8\Delta_{us}^2\left(\Delta_{ud}^2+4R^2\right)
+\left(\Delta_{ud}^2-4R(\epsilon_p+R)\right)^2\right)^{1/2}.
\ee
The result becomes
easier to understand if we consider some simple limits. If we ignore
flavor symmetry breaking, $R=0$, and set $\Delta_{ud}=\Delta_{us}$
we find 8 eigenvalues $p_0\pm(\epsilon_p^2+\Delta^2)^{1/2}$ and one
eigenvalue with the gap $2\Delta$. This is indeed the expected spectrum
in the color-flavor locked phase. If, on the other hand, we set
$\Delta_{us}=0$ we find 4 eigenvalues $p_0\pm(\epsilon_p^2+\Delta^2
)^{1/2}$ while the other 5 eigenvalues have vanishing gaps. Again,
this is as expected.

  We note that in the presence of flavor symmetry breaking the first
three eigenvalues, which depend on $\Delta_{ud}$ only, are completely
unaffected. For the next 4 eigenvalues, which only depend on
$\Delta_{us}$, the energy $p_0$ is effectively shifted by $R$. This
is exactly as in the simple toy model discussed above. It implies that
for $R>\Delta_{us}$, when we close the integration contour in the
complex $p_0$ plane, we do not pick up this pole. The last two
eigenvalues are more complicated. They depend on both $\Delta_{ud}$
and $\Delta_{us}$, and they explicitly contain the flavor symmetry
breaking parameter $R$. Nevertheless, the structure of the eigenvalues
is certainly suggestive of the idea that for $R<\Delta_{us}^0$ we
have $\Delta_{us}\simeq \Delta_{ud}$, and the gaps are almost
independent of $R$, while at $R\simeq\Delta_{us}^0$ there is a
discontinuity and $\Delta_{us}$ goes to zero.

This is borne out by a more detailed calculation. For this purpose,
we add a flavor and color anti-symmetric short range interaction
\be
{\cal L} = \frac{K}{4}\big( \delta^{\alpha\gamma}\delta^{\beta\delta}
 -\delta^{\alpha\delta}\delta^{\beta\gamma} \big)
 \big( \delta_{ac}\delta_{bd}-\delta_{ad}\delta_{bc}\big)\,
 \left(\psi^\alpha_a\sigma_2\psi^\beta_b\right)
 \left(\psi^{\gamma\,\dagger}_c\sigma_2 \psi^{\delta\,\dagger}_d\right).
\ee
The free energy of the system is the sum of the quasi-particle
contribution from (\ref{evals}) and the mean field potential
$V=1/K\cdot(\Delta_{ud}^2+2\Delta_{us}^2)$. There are two coupled
gap equations, which can be derived by varying the free energy
with respect to the two parameters $\Delta_{ud}$ and $\Delta_{us}$.
Numerically, however, it is much simpler to minimize the free
energy directly. As an example we show results for $\mu=0.5$
GeV and $\Delta_{us}^0=25$ MeV in Figure 1. We observe that
the flavor symmetry breaking difference $\Delta_{ud}-\Delta_{us}$
is very small all the way up to the critical strange quark mass.
At the critical mass, there is a discontinuous transition to a phase
where $\Delta_{us}$ vanishes exactly. The value of the critical
mass is very close to the estimate $m_s=2\sqrt{\mu\Delta_{us}^0}$.
We should note that in the case of color-flavor locking there
are two physical gap parameters, the octet and the singlet gap.
The scale for the critical strange quark mass is set by the
octet gap.

\section{Two Flavors}

When $m_s$ is effectively large according to the preceding criterion,
it becomes legitimate, when analyzing the light two flavors, to neglect
the influence of the strange quark. This implies that the two-flavor
analysis carried out in \cite{ARW_98,RSSV_98} is appropriate. The
major result of that analysis is the existence of a condensate of
the form
\be
\label{del_s}
\langle q_{L\,a}^{\alpha}C q_{L\,b}^{\beta}\rangle =
-\langle q_{R\,a}^{\alpha}C q_{R\,b}^{\beta}\rangle =
\Delta \epsilon^{\alpha\beta 3}\epsilon_{ab}.
\ee
This condensate imparts a very substantial gap, plausibly of order 100
MeV at several times nuclear density, to two of the quark colors. It
leaves an $SU(2)$ subgroup of color $SU(3)$, acting among these two
colors of quarks, unbroken. The third color of quark does not acquire
a gap from the primary condensate. As we will discuss below, these
quarks can acquire a small gap from secondary modes of condensation.
This gap, however, will not break the residual gauge symmetry, since
it only involves quarks of the third color. The residual gauge
symmetry can be broken in the case of three flavors, even if
the strange quark is heavy, as long as  the chemical potential
is above the onset for strangeness. We will discuss possible
modes of strangeness condensation below.

The expected low energy degrees of freedom are quasiparticle
excitations around the Fermi surfaces of the third
color quarks. It is important to note that in these channels, there
are no massless gluons. These low energy degrees of freedom are truly
weakly coupled. Gluons in the unbroken $SU(2)$ remain
massless, but their interactions become strong.  We expect this means that
the $SU(2)$ is confined, just as it would be in a vacuum theory
with massive quarks.  Of course, here we cannot claim rigorous control.

The primary condensate (\ref{del_s}) leaves a residual gauge and
chiral symmetry, as well as an unbroken $U(1)$ baryon number
symmetry. This can be seen as follows: The condensate breaks
both color hypercharge $Q_8={\rm diag}(1/3,1/3,-2/3)$ and baryon
number $B={\rm diag}(1/3,1/3)$, but leaves the combination
$B'=B-Q_8$ invariant. Similarly, the condensate violates
electric charge $Q_{em}={\rm diag}(2/3,-1/3)$, but there
is a modified charge operator $Q'_{em}=Q_{em}-Q_8$ which
leaves the condensate invariant. So despite initial appearances the
primary condensate is not an electric superconductor.  A modified
photon remains massless.

Under the unbroken symmetries, the quarks of the third color
carry baryon number one.  In addition to that,
they are singlets under the residual $SU(2)$ gauge symmetry. And they
carry charges (1,0) (in electron charge units)
with respect to the modified photon.  Thus,
at asymptotically large density, the low energy -- quark -- degrees
of freedom have the quantum numbers of the proton and the neutron!

At the level of the primary condensate,
there is no gap for these modes. It is interesting
to note, however, that the possible modes of secondary condensation
involving these modes correspond
exactly to the known possibilities for pairing in nuclear matter.
This is the case because the remaining quarks not only carry the
same isospin and charge quantum numbers as the proton and neutron
but also, as emphasized above, they are both singlets under the residual
gauge group.
The superfluid pairings most often considered in the context
of nuclear matter have the quantum numbers $^{2s+1}L_J={^1S_0}$, $^3S_1$
and $^3P_2$. While $^1S_0$ is expected to dominate at low density,
$^3P_2$ pairing is likely to take over at high density. The
simplest operators with these quantum numbers are
\bea
^1S_0 &\hspace{1cm}&  \psi C\gamma_5\tau_2\vec\tau\psi \\
^3S_1 &\hspace{1cm}&  \psi C\gamma_5\vec\gamma\tau_2\vec\tau\psi,
       \hspace{0.3cm} \psi C\vec\Sigma\tau_2\psi \\
^3P_2 &\hspace{1cm}&  S\psi C\gamma_5\gamma_i\hat k_j\tau_2\psi, \ldots
\eea
Here, $\hat k=\vec{k}/|\vec{k}|$ is the unit momentum vector
and $S$ projects out the symmetric traceless part. Nothing
essentially new is gained by considering operators that involve
the coupling of spin and angular momentum, such as $\psi C\vec
\gamma\hat k\psi$. These operators can be reduced using the
equations of motion.

One gluon exchange is repulsive in all these channels, with
the exception of the tensor operator $\psi C\vec\Sigma\tau_2\psi$.
That is the channel that was considered in \cite{ARW_98}. In
this work, the gap in the tensor channel was estimated to be in
the range of several up to 100 keV.  Of course, this estimate is
exponentially sensitive to poorly determined couplings.

For later purposes, let us note that each of these possible secondary
condensates breaks
either isospin or rotational symmetry.

It is remarkable that in
a theory where the dominant scale, the gap of the first two
color quarks, is on the order of 100 MeV, the mass gap of the
light degrees of freedom comes out to be so small. This is
clearly reminiscent of the situation in real-world QCD at small density.
QCD dynamically generates masses on the order of several
hundred MeV, but the binding energy of nuclear matter and the
gap in superfluid nuclear matter are small, on the order of
a few MeV.   This disparity is among the most fundamental
qualitative facts about nuclear physics that one would like to understand
from a microscopic viewpoint.  From all previous standpoints known to
us, it results from a conspiracy.  Since just such a disparity
appears as a calculable feature of
the high-density phase, we are motivated to consider whether
it is legitimate to extrapolate from the
high-density phase to nuclear density.  If it is, then
no conspiracy need be invoked.

To justify such extrapolation, however, we must assume that there
is no significant phase transition separating the high-density color
superconducting phase
from nuclear matter.  (Changes in the nature of the small secondary
condensate, which  are certainly
expected in view of the discussion above, are not germane here.)
This has the startling implication, that chiral symmetry must be  -- up to
the effects of non-zero light quark masses, electromagnetism, and
small secondary
condensations -- restored in nuclear matter.  Whether this is true in
reality, is worthy of much further investigation.   It is suggestive,
in this regard, that
several experimental determinations suggest that the axial vector coupling
$g_A$, whose free-nucleon value 1.26 reflects chiral symmetry breaking
\cite{adler-weisberger}, is renormalized to very nearly unity in
nuclear matter \cite{buck} \cite{brown} \cite{ericson-weise}.

\section{One Flavor}

Above the onset for net strangeness, there is a Fermi surface
for strange quarks. For large $m_s$ there will be a big Fermi surface
mismatch, and the strange quark will not
pair with light
quarks, as we discussed earlier.  Then we can analyze its condensation
independently.
We expect a purely strange diquark
condensate to form. This condensate breaks strangeness, but
conserves a $Z_2$ that flips the sign of the strange quark.

Possible modes of strange quark superfluidity were studied
in detail by Bailin and Love \cite{BL_84}. If the condensate
is to be color antisymmetric, overall symmetry requires that
it cannot be a scalar. A particularly interesting possibility
is to lock color and spin
\be
  \langle s^\alpha C\gamma_i s^\beta\rangle =
    \Delta \epsilon^{\alpha\beta\gamma}\delta^\gamma_i .
\ee
With this condensate, color symmetry is completely broken, as is
naive rotational symmetry, but a modified rotation symmetry involving
simultaneous color and naive spatial rotation remains valid.

By taking the
non-relativistic limit, and ignoring the complications due to
antiparticles, we can vastly simplify the analysis of this
phase. We consider the action
\be
\left(\psi^\dagger\hspace{0.3cm} \psi\right)
\left(\begin{array}{cc}
p_0-\epsilon_p    & \Delta_{csl} M_{csl} \\
\Delta^*_{csl} M_{csl}^*& p_0+\epsilon_p
\end{array}\right)
\left( \begin{array}{c}
\psi \\ \psi^\dagger
\end{array}\right) ,
\label{gor_csl}
\ee
where $\psi$ is now a six-dimensional spin-color spinor. The
pairing matrix $M_{csl}$ corresponding to color-spin locking
is given by
\be
M_{csl} = (\sigma_2\sigma_k)_{ij}\epsilon^{k\alpha\beta}.
\ee
Other possible forms of order are
\be
\begin{array}{lcll}
M_{pol} &=& (\sigma_2\sigma_3)_{ij}\epsilon^{3\alpha\beta}
& {\rm polar} \\
M_{A} &=&  (\sigma_2\sigma_+)_{ij}\epsilon^{3\alpha\beta}
& {\rm A-phase} \\
M_{pla} &=& (\sigma_2\sigma_1)_{ij}\epsilon^{1\alpha\beta}+
(\sigma_2\sigma_2)_{ij}\epsilon^{2\alpha\beta} & {\rm planar}
\end{array}
\ee
Let us first look at the color-spin locked phase. The quadratic
action (\ref{gor_csl}) is easily diagonalized. We find four
eigenvalues with gap $\sqrt{2}\Delta_{csl}$ and two eigenvalues
with gap $\Delta_{csl}$. The degeneracies reflect the unbroken
rotational symmetry. The state with gap $\sqrt{2}\Delta_{csl}$
has spin 3/2 and degeneracy $2s+1=4$, while the state with gap
$\Delta_{csl}$ has spin 1/2 and degeneracy $2s+1=2$. In the polar
phase, there are four states with gap $\Delta_{pol}$ while two
states remain gapless. In the A-phase, only two quarks acquire
a gap $\Delta_A$, while the remaining four are gapless. In the
planar phase, finally, all the physical excitations are gapless.

Since the color-spin locked phase is the only phase where
all quarks acquire a gap it seems reasonable to expect that
this is the groundstate of one flavor QCD at zero temperature
and large chemical potential. This is what Bailin
and Love found by analysing the Landau-Ginzburg functional, which
however is only justified in the
vicinity of the critical temperature.  
For simple model interactions, we find
that either color-spin locking or the polar phase might
be favored at zero temperature.  A rigorous treatment is a feasible and
interesting project, but will not be attempted here.

It is amusing to note that the color-spin locked phase has
excitations with spin 3/2. This, of course, is the spin of the
baryon in QCD with one flavor. The ordering of the states,
however, appears to be inverted as compared to
the expectation from the hadronic phase: while the ground state
baryon has spin 3/2 in the hadronic phase, and the spin 1/2 baryon
is an excited state, a spin 1/2 excitation is lowest the two in the
high density phase.  This might indicate that the high-density and
nuclear phases must be separated by a phase transition, though
strictly speaking it need not, since the ordering of low-lying
levels is a non-universal feature.  In any case, the situation is
certainly less straightforward than for 
three degenerate
flavors.

\section{Two Plus One Flavors}

We have suggested the existence of three major types of
ordered phases for high-density QCD with 2+1 flavors: two-flavor color
superconductivity with or without an accompanying strange superfluid, 
and color-flavor locking.
Our description of each differs substantially, and our modelling
above (and immediately below) suggests they are separated by sharp
phase transitions, but the question arises whether
the need for such transitions is guaranteed by any mismatch of
physical symmetry.  (Differences in gauge symmetry, which are
meaningful only in the context of weak coupling,  cannot be invoked
here \cite{R_75} \cite{FS_79}.)

Fortunately, it is.  The two-flavor
color superconducting state, given only the primary condensation,
has the global symmetry
$SU(2)_L \times SU(2)_R \times U_S (1) \times {\tilde U}_B(1)$ for
massless light quarks, where the first two factors are chiral isospin,
the third is ordinary strangeness, and the fourth is (modified) baryon
number.   With non-zero but degenerate light quark masses, the first
two factors collapse into a diagonal $SU(2)$.  The strange
superconducting state differs from this by breaking
$U_S(1) \rightarrow Z_2$.  The color-flavor locked state has $\tilde
{SU}(2) \times {\tilde U} (1) $.  These clearly all differ.  However
in the two-flavor superconducting state, since the primary
condensation leaves gapless modes,  we must in addition consider the
effect of possible secondary condensations.  However, as we already
mentioned in passing, these condensations break either isospin or
rotational symmetry (but not strangeness), so they cannot reproduce
the symmetry of either of the other two major phases.

Our discussion of the phase structure of dense matter in QCD with
two light and one intermediate mass flavor can be illustrated
concretely with the help of a simple schematic model. Let us
consider a short range interaction with the quantum numbers of one
gluon exchange
\be
  {\cal L} = K(\bar\psi\gamma_\mu\lambda^a\psi)
 (\bar\psi\gamma^\mu\lambda^a\psi),
\ee
characterized by a coupling constant $K$. We study the phase
structure of the system in the mean field approximation. Since
we are interested in chiral symmetry breaking and dynamical
mass generation as well as superfluidity we can no longer
restrict ourselves to particle-hole pairs, but have to
include anti-particle (hole) contributions also. The free
energy is of the form
\bea
 F &=& \sum_i \epsilon(\sigma_i,\delta_i) + V ,\\
 \epsilon(\sigma,\delta) &=& \int\frac{d^3p}{(2\pi)^3}\left(
 \sqrt{(E_p-\mu)^2+\delta^2}+\sqrt{(E_p+\mu)^2+\delta^2}\right),
\eea
where $E_p=\sqrt{p^2+(\sigma+m)^2}$ is the single particle
dispersion relation and $V$ is the mean field potential.
The integration over $p$ is regularized by a sharp cutoff
$\Lambda$.

We consider the following phases:

1) A phase with chiral symmetry breaking only. This phase is
characterized by a chiral condensate
\be
 \langle\bar q^\alpha_a q^\beta_b\rangle =
 \delta^{\alpha\beta} \big( \delta_{ab} \Sigma_0 +
 \delta_{a3}\delta_{b3} \Sigma_s \big).
\ee

2) The color flavor locked phase, including the possibility
of a dynamically generated contribution to the strange quark
mass
\bea
 \langle q^\alpha_a C\gamma_5 q^\beta_b\rangle &=&
 \big( \delta^\alpha_a\delta^\beta_b \Delta_1 +
 \delta^\alpha_b\delta^\beta_a \Delta_2 \big),\\
 \langle\bar s^\alpha s^\beta\rangle \;\; &=&
 \delta^{\alpha\beta}\Sigma_s
\eea

3) The superconducting phase, with or without strange
quark condensation
\bea
 \langle q_{a}^{\alpha}C\gamma_5 q_{b}^{\beta}\rangle &=&
 \epsilon^{\alpha\beta 3}\epsilon_{ab3}\Delta_{ud} \\
 \langle\bar s^\alpha s^\beta\rangle \;\;&=&
 \delta^{\alpha\beta} \Sigma_s.
\eea
In practice we ignore the condensation energy of the
light quarks of the third color and the strange quark
superfluid, since both of them are expected to be
quite small.

The effective potential and the mass gaps in the three
different phases are easily calculated. We find
\bea
1)    \hspace{1cm}
\sigma_1 &=& \ldots =\sigma_6 = \frac{16}{3}K\Sigma_0 \nonumber\\
\sigma_7 &=&\ldots  =\sigma_9 = \frac{16}{3}K(\Sigma_0+\Sigma_s)
\nonumber\\
     V   &=& K\left(48\Sigma_0^2 + 32\Sigma_0\Sigma_s
              + 16 \Sigma_s^2 \right)  \nonumber\\
2)   \hspace{1cm}
\delta_1 &=& \ldots =\delta_8 = \frac{K}{3} \left(3\Delta_1
 -\Delta_2 \right) \hspace{0.5cm}
\delta_9  =  \frac{K}{3} 8\Delta_2 \nonumber\\
\sigma_7 &=& \ldots =\sigma_9= \frac{16}{3}K\Sigma_s \\
     V   &=& K\left(-16 \Delta_1\Delta_2+16\Sigma_s^2\right) \nonumber\\
3)  \hspace{1cm}
\delta_1 &=& \ldots  =\delta_4 = \frac{4}{3}K\Delta_{ud}\nonumber\\
\sigma_7 &=& \ldots  =\sigma_9 = \frac{16}{3}K\Sigma_s  \nonumber\\
V   &=& K\left(\frac{16}{3}\Delta_{ud}^2+16\Sigma_s^2\right)
\nonumber
\eea
In the color-flavor locked phase we include the shift in the
eigenvalues due to the strange quark mass, following our
detailed discussion in the first section. In particular, the integral in
the non-strange-strange particle-hole channel is restricted to
$\bar\epsilon_p^2>R^2+\delta^2$. The only difference is that
we allow for the possibility of a dynamically generated
contribution to the strange quark mass. It is now straightforward
to minimize the free energy in the three phases separately and to
determine the global minimum by comparing the different solutions.
The result is shown in Figure 2. We have chosen $\Lambda=600$ MeV and
fixed $K=32\Lambda^{-2}$ in order to reproduce a $\mu=0$ quark
constituent mass of 400 MeV.

We note that each of the phases disussed above is indeed favorable for
some values of the chemical potential and strange quark mass.
The transition from the color-flavor locked phase to the nuclear
or hypernuclear phase is shifted to smaller strange quark masses
as compared to the calculation in section II. This effect is due
to the large dynamically generated strange quark mass. Also, as a function
of the chemical potential, there is an intermediate nuclear phase
between the vacuum and color-flavor locked phases even for very
small strange quark mass. Again, this is due to the large dynamically
generated strange quark mass. We also note that even for small
strange quark masses $m_s\sim 50$ MeV, the onset for strangeness
is significantly above the onset for light quarks.  This, of course,
is the most direct manifestation of a dynamically generated strange
quark mass.

We should stress that the model is clearly very crude, and in
particular that
the locations of the phase boundaries should not be regarded as
quantitative predictions.  Nevertheless, we believe that the existence
of the different phases, and the topology of the phase diagram
are robust features consequences of the physical mechanisms
we have discussed.

There is an interesting consequence, in the strange superconductor phase,
of the interplay between
color-spin locking for the strange quarks and color symmetry breaking
induced by the primary light quark condensation.  The preferred color
axis induces a preferred spin axis, so that the material becomes a
strange ferromagnet, in particular violating even the modified
rotation symmetry.

Finally, we would like briefly to discuss two interesting
possibilities for high density that have been proposed and much
debated in the literature, from the point of view ({\it i.e}.,
working down from
the asymptotics) used here.   One is the hypothesis of stable strange
quark matter \cite{Wit_84}.  In our language, it is the hypothesis
that in the interacting theory there is no separate onset transition
for strangeness, but strange particles appear immediately and
abundantly at the
``void to matter'' transition.  In our Figure 2, it would mean that
the lower portion of the sc region would be squeezed out.   Although
in the nature of things we cannot rule this out, since it is proposed
as a strong-coupling phenomenon and our methods are intrinsically
weak-coupling, nothing we have encountered suggests it.   The second
is K-condensation \cite{kcond}.  It is of course
perfectly trivial to write
down bilinear quark-antiquark condensates that match the hadronic
K-condensation quantum numbers (namely, $\langle \bar s \gamma_5 u
\rangle$ or $\langle \bar s \gamma_0 \gamma_5 u \rangle$).  Only
slightly less trivial is that one can do so with the residual ungapped
degrees of freedom, after the primary condensation in two-flavor
superconductivity.  Again, however, nothing we have encountered
suggests this is a very favorable possibility. 

\centerline{\bf Acknowledgement}
 Related issues are addressed in the paper ``Unlocking
Color and Flavor in Superconducting Strange Quark Matter'', by
M. Alford, J. Berges, and K. Rajagopal, MIT preprint MIT-CTP-2844,
appearing simultaneously.   Where they overlap, our conclusions
agree.  We thank these authors for showing us their work prior to
publication, and for informative discussions.


\newpage\noindent
\begin{figure}
\caption{\label{fig_cfl}
Light quark gap $\Delta_{ud}$ and strange-non-strange gap $\Delta_{us}$
as a function of the strange quark mass. }
\end{figure}

\begin{figure}
\caption{\label{fig_phase}
Phase diagram calculated in the schematic model described in Section IV.
We show the phase boundaries between the vacuum phase (void), the
color-flavor locked phase (cfl), the nuclear phase (sc), and the
hypernuclear phase (scs). Note that the transition line from the nuclear
phase to the color-flavor locked phase turns sharply towards the
void-cfl transition at very small $m_s$. For $m_s=0$ there is a
direct transition from the vacuum phase to the color-flavor locked 
phase. Also note that the phase diagram was calculated with a 
$\Lambda=600$ MeV form factor, so the turnover of the cfl-scs
transition near $\mu=600$ MeV is an artefact of this formfactor.}
\end{figure}

\setcounter{figure}{0}

\begin{figure}
\begin{center}
\vspace{4cm}
\epsfxsize=16cm
\epsffile{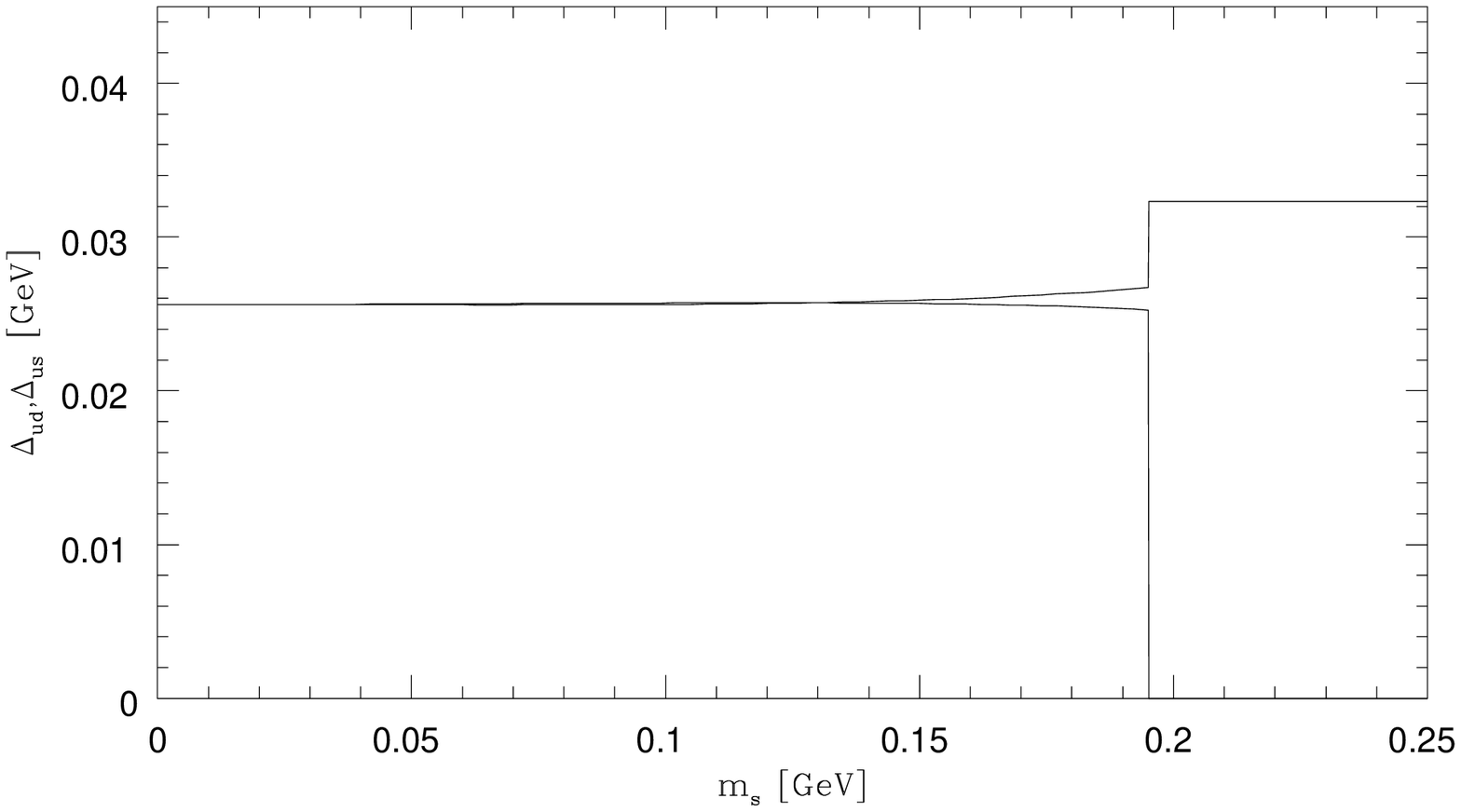}
\end{center}
\caption{}
\end{figure}

\begin{figure}
\begin{center}
\epsfxsize=16cm
\epsffile{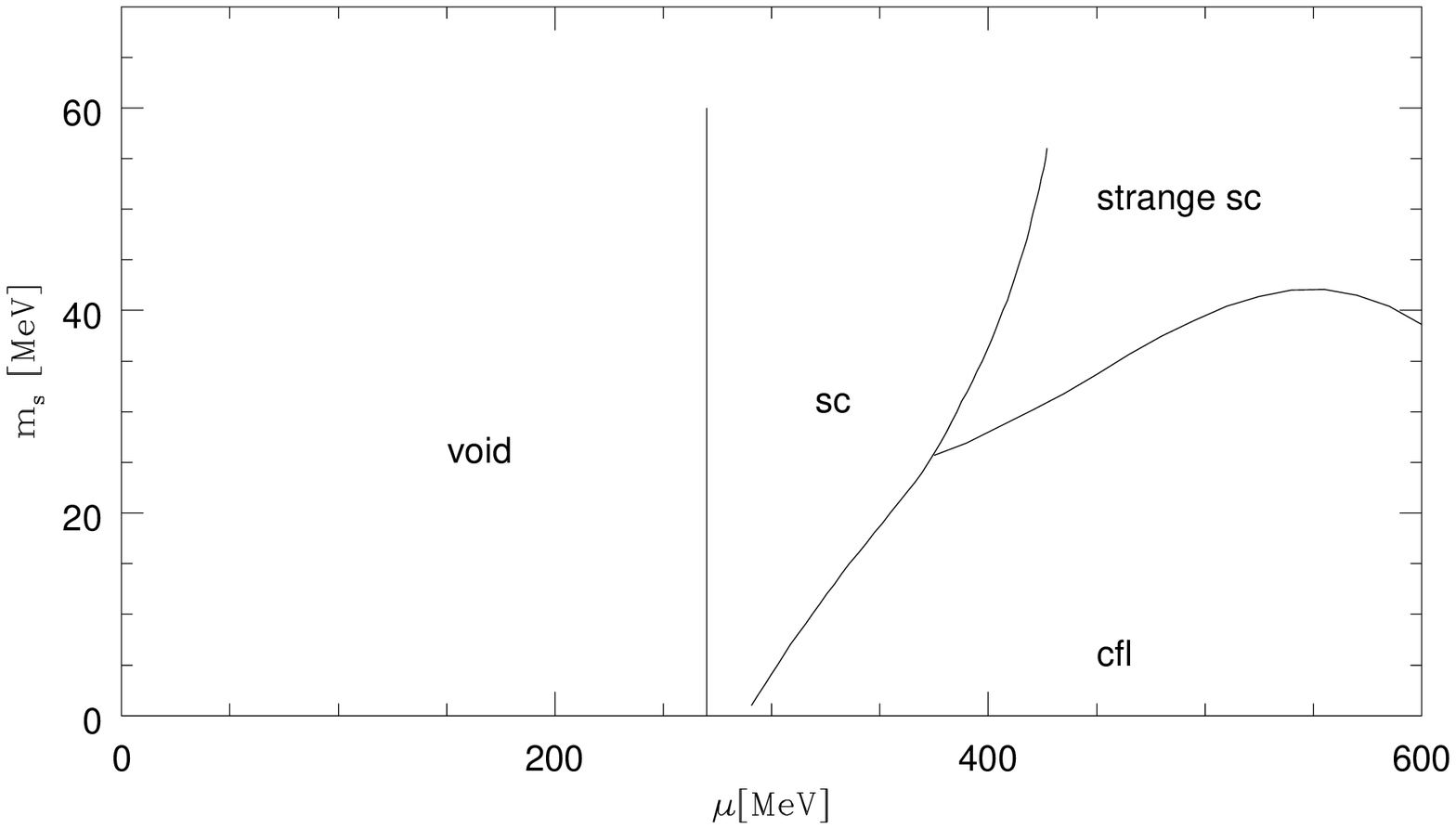}
\end{center}
\caption{}
\end{figure}

\end{document}